\documentclass[journal]{IEEEtran}

\ifCLASSINFOpdf
  \usepackage[pdftex]{graphicx}
  \DeclareGraphicsExtensions{.pdf,.jpeg,.png}
\else
  \usepackage[dvips]{graphicx}
  \DeclareGraphicsExtensions{.eps}
\fi
\usepackage[cmex10]{amsmath}
\usepackage{amsthm}
\usepackage{amssymb}
\theoremstyle{plain}

\newtheorem{thm}{Theorem}

\theoremstyle{definition}

\begin{document}

\title{Multipath-dominant, pulsed doppler analysis of rotating blades}

\author{Michael~Robinson$^{(1)}$
\thanks{(1) Department of Mathematics, University of Pennsylvania,
209 S. 33rd Street, Philadelphia, PA 19104 USA e-mail:
robim@math.upenn.edu}}% <-this % stops a space
\markboth{}{Multipath analysis of rotating blades}


\maketitle

\begin{abstract}
We present a novel angular fingerprinting algorithm for detecting changes in the direction of rotation of a target with a monostatic, stationary sonar platform.  Unlike other approaches, we assume that the target's centroid is stationary, and exploit doppler multipath signals to resolve the otherwise unavoidable ambiguities that arise.  Since the algorithm is based on an underlying differential topological theory, it is highly robust to distortions in the collected data.  We demonstrate performance of this algorithm experimentally, by exhibiting a pulsed doppler sonar collection system that runs on a smartphone.  The performance of this system is sufficiently good to both detect changes in target rotation direction using angular fingerprints, and also to form high-resolution inverse synthetic aperature images of the target.
\end{abstract}


\section{Introduction}
\IEEEPARstart{T}{he} doppler structure of rotating targets is of considerable recent interest.  Most of this effort has focused on detection and classification, often with an eye toward filtering and removal.  For instance, the proliferation of wind turbines presents challenging doppler clutter \cite{Poupart_2003} for air traffic control radars.  Many methods exploit micro-doppler structure; specifically those portions of a target's echo that are localized in range and doppler near the target echo's centroid.  In favorable settings, micro-doppler structure contains sufficient information to form an inverse synthetic aperture image of a target.  However, if the sensor platform is monostatic and the target's centroid is fixed, there remains an unavoidable ambiguity: the direction of rotation cannot be determined.

This ambiguity is resolved if one extends treatment to echos outside the immediate vicinity of the target's range and doppler, and considers {\it multipath} signals.  Unfortunately, methods for processing these signals are lacking, and usually require intricate knowledge of the environment.  For instance the algorithms presented in \cite{Tayebi_2009} require the construction of a detailed ray-tracing model of the environment.  We present a methodology for processing doppler multipath of rotating targets that makes only limited assumptions about the environment, since it is based on the theory of differential topology.  We rely on the framework introduced in \cite{GhristRobinson}, which presented a unified method for detection and localization under general settings.  We validate this topological methodology in simulation and with a controlled set of experiments using indoor sonar returns of a rotating target.

We present a novel {\it angular fingerprinting} algorithm, which allows changes in direction of rotation to be detected.  Since the algorithm is based on the inherent topological robustness of multipath doppler signatures, it is unnecessary to use a separate detection process to identify which portions of the signal constitute multipath echos.  As a convenient aside, our algorithm can be demonstrated easily with low-quality equipment.  We therefore also present a feasibility study for sonar imaging and acoustic multipath measurement using consumer-grade audio equipment.  We focus on the common ceiling fan as our target, using it as a proxy for more complicated rotating targets.

\section{Historical treatment}
With the growth of the wind power industry, there has been increasing concern about the doppler clutter they present to tracking radar systems.  Perhaps the earliest detailed account of the effects of wind farms on radar systems was given in \cite{Poupart_2003}.  In that report, both modeling and controlled observations of typical wind turbines were described.  They concluded that there is substantial structure in the echos in the immediate vicinity of a wind turbine (microdoppler) as well as multipath structure that extends farther away from the turbine's range.  In the following years, various {\it ad hoc} techniques have been devised (such as \cite{Perry_2007}) to exise these effects from radar data.  In an effort to be more systematic, \cite{Isom_2009} performed high-fidelity measurements of wind turbines to characterize their microdoppler structure.  

With the availability of high-resolution inverse synthetic aperature systems, microdoppler structure has become a valuable source of information about moving targets, including wind turbines.  Chen and his colleagues \cite{Chen_2001,chen_2006} have performed detailed simulations to classify microdoppler effects.  Thayaparan and his colleagues \cite{Thayaparan_2006,Thayaparan_2007} further showed that rotating portions of targets can be extracted from time-frequency and wavelet analysis of their microdoppler structure.  In particular, this enables one to separate rotating, static, and linear trajectories from each other.  Assuming that the target centroid is in motion, one has nonzero crossrange resolution and can then solve for the direction of rotation.  Multistatic systems \cite{Suwa_2009} can provide an instantaneous rotation vector, if there is relative motion of centroids.  In contrast with this approach, we show that {\it changes} in rotation direction can be detected with a monostatic system even if there is {\it no} crossrange resolution by using multipath signals instead.  

Multipath has long been known to contain potentially valuable, if difficult to extract, information for target localization.  For instance, Sherman \cite{Sherman_1971} showed that multipath echos collected from a single pulse could be used to provide a localization solution of a moving target.  Extracting these responses from the background can be complicated; the interested reader can find a systematic treatment in \cite{Hahm_1997}.  However, since this kind of systematic treatment is often more complicated than is desirable for real-time systems, the simpler method of {\it multipath fingerprinting} has become extremely common. (The interested reader should consult \cite{Kelly_2000}, \cite{Prasith_2002}, \cite{Nerguizian_2006}, \cite{Roxin_2007}, \cite{Fang_2008}, or \cite{Tayebi_2009} for representative treatments.)  Fingerprinting methods rely on the fact that the received multipath structure is uniquely determined by the location of a receiver in an environment illuminated by a collection of transmitters whose locations may be unknown.  In order to provide useful results, fingerprinting methods require the collection of a reference dataset, in which known locations are associated with the multipath response at that location.  Multipath fingerprints are often obtained opportunistically, using an existing communication infrastructure as a network of illuminators.

The setup of multipath fingerprinting generalizes considerably as shown in \cite{GhristRobinson}, and can be used to localize targets or changes in the environment.  In this article, we apply a simplified fingerprint approach that differs from the usual ones.  Like the usual approaches, we require a reference collection to be acquired.  However, instead of using a passive approach, we actively sense the environment by introducing a waveform with fine range resolution and doppler sensitivity.  Because of this, we implicitly include doppler information in the resulting fingerprints.  Usual multipath fingerprinting requires a sophisticated fingerprint matching algorithm because the resulting data is high-dimensional.  In contrast, our space of locations to be sensed is one-dimensional, and therefore only a few multipath signals are necessary.  Further, because our rotating target has substantial angular momentum, we can limit the complexity of the matching algorithm to a simple Kalman filter. 

Since it is our desire to work with simple algorithms and equipment, we demonstrate performance of our angular fingerprint algorithm on a system built from consumer-grade equipment.  Surprisingly few other researchers have made use of such hardware; we want to advocate for its cost-effectiveness for rapid prototyping.  Perhaps the best {\it tour de force} in this area is Charvat's synthetic aperature radar, powered by a garage door sensor. \cite{Charvat_2006}  Like the sonar measurements we take, sonar measurements have been demonstrated by others using a PC sound card \cite{Matejowsky_2008}, an iPhone \cite{Laan_2009}, or an Android smartphone \cite{Dicon_2012}.  Indeed, \cite{Matejowsky_2008} successfully demonstrated low-resolution synthetic aperature image formation.  However, we believe our article to be the first use of a smartphone's sound card to form an {\it inverse} synthetic aperature image of a moving target.

\section{Theoretical discussion}

Consider a propagation environment with a single rotating, reflective target.  For concreteness, the simulation and experiment presented in this article focuses on a potentially complicated room with a fan spinning at a fixed, known location.  Formally, the entire configuration of the environment and target is parametrized by a single circle-valued coordinate.

Suppose a sensor has been placed in the environment that instantaneously takes a collection of $N$ real-valued measurements (which might be signal levels at a particular time, times of arrival, or other similar measurements).  In this case, we have a function $S^1 \to \mathbb{R}^N$ taking the angular coordinate of the target to the recieved measurements.  Notice that this function {\it implicitly} encodes sensor location(s), sensor modality, and many processing considerations.  In this article, the measurements will be either the times of arrival of the first $N$ echos or the signal levels at a collection of $N$ time samples within each pulse interval.

We can generalize this slightly to permit measurements to fail, which can happen if an echo is shadowed.  In this case, we permit measurements to take the value $\perp$ to indicate a non-$\mathbb{R}$ value.  In this case, we expand our function to be $P:S^1\to(\mathbb{R}\,\sqcup\perp)^N$.  Following \cite{GhristRobinson}, we call such a function $P$ a {\it signal profile} when certain regularity conditions are met.  Specifically, if there exists a cover $\mathcal{U}=\{U_1,...,U_M\}$ of $S^1$ so that 
\begin{enumerate}
\item each $U_i$ is a compact, 1-dimensional set,
\item every point in the image $P(U_i)$ has the same set of non-$\perp$ coordinates, (we write $N_i$ for the number of non-$\perp$ coordinates in $P(U_i)$) and
\item the projection $P_i:U_i \to \mathbb{R}^{N_i}$ to the non-$\perp$ coordinates of the restriction $P|U_i:U_i\to(\mathbb{R}\,\sqcup\perp)^N$ is a smooth function.
\end{enumerate}

The principal result about signal profiles is the following
\begin{thm}
\label{inj_thm}(Immediate consequence of \cite{GhristRobinson}, Theorem 3, the Signals Embedding theorem)
The signal profile $P:S^1\to(\mathbb{R}\,\sqcup\perp)^N$ arising from a generic set of maps $P_i \in C^\infty(U_i,\mathbb{R}^{N_i})$, is injective when the minimal $N_i$ is greater than 3.
\end{thm}
Intuitively, this means that signal profiles in practice will usually be injective if enough measurements are taken at each configuration, and (by the inverse function theorem) will have locally-defined smooth inverses.

In the case of a single rotating target, this means that the angular position of a rotating target is completely determined if there are always at least three independent measurements available for each angle.  This is robust to multipath: often particular reflections will only be visible for a particular range of angles before being shadowed.

Ambiguities can arise if too few measurements are made.  For instance, if the sensor measures only the shortest path distance to a reflection point on the target, there is an unavoidable symmetry: all points that have the same off-boresight angle are indistinguishable.  In essence, Theorem \ref{inj_thm} suggests that by collecting multipath signals, this ambiguity is removed.

Of course there does remain a particular ambiguity in the case of a rotating fan; if there are $K>1$ blades, the image of the signal profile will be more highly periodic, namely $P(\theta+2\pi/K)=P(\theta)$.  Indeed, a truly {\it generic} perturbation of $P$ would destroy this periodicity and destroy the ambiguity, possibly by changing the effective lengths of the blades.  However, we will not proceed in this direction in this article as this is unrealistic.

This {\it ansatz} suggests a methodology (see Section \ref{finger_alg_sec} for details) for the measurement of rotation direction and relative speed: first acquire one rotation of the target, storing the measurements taken.  Later acquisitions can be compared to these previous measurements to recover the angular position of the target by simply finding the nearest match using a metric on $(\mathbb{R}\,\sqcup\perp)^N$.

\section{Experiment design and data processing}
Theorem \ref{inj_thm} makes a strong statement about the ability to detect changes in the direction of rotation of targets and to measure their rotation rate.  However, its reliance on multipath signals means that it may be difficult to exploit in practice due to limitations imposed by noise and clutter.  Additionally, it is possible that hidden symmetries imposed by sampling or experimental structure could prevent the typical experiment from having sufficiently many {\it independent} signals.  Therefore, we conducted experiments both in simulation and with simple acoustic equipment to verify that direction and rate of rotation are detectible in practice.  In so doing, we also demonstrate the power of consumer-grade acoustic hardware for the purpose of synthetic aperture imaging.

Specifically, the experiments aim to verify the following properties of the signal profile:
\begin{enumerate}
\item Repeatability: if the system returns to the same physical state (angle of rotating target), how similar are the associated multipath signals?
\item Injectivity: are there any distant pairs of angles of the target that result in the same or similar multipath?
\end{enumerate}

\subsection{Experimental setup: simulation and physical}
We conducted two kinds of experiments, in simulation and physically.  The simulation provides relatively pristine, easy to interpret results.  On the other hand, conclusive success of the physical experiments demonstrates the robustness of our approach.

\begin{figure}
\begin{center}
\includegraphics[width=1in]{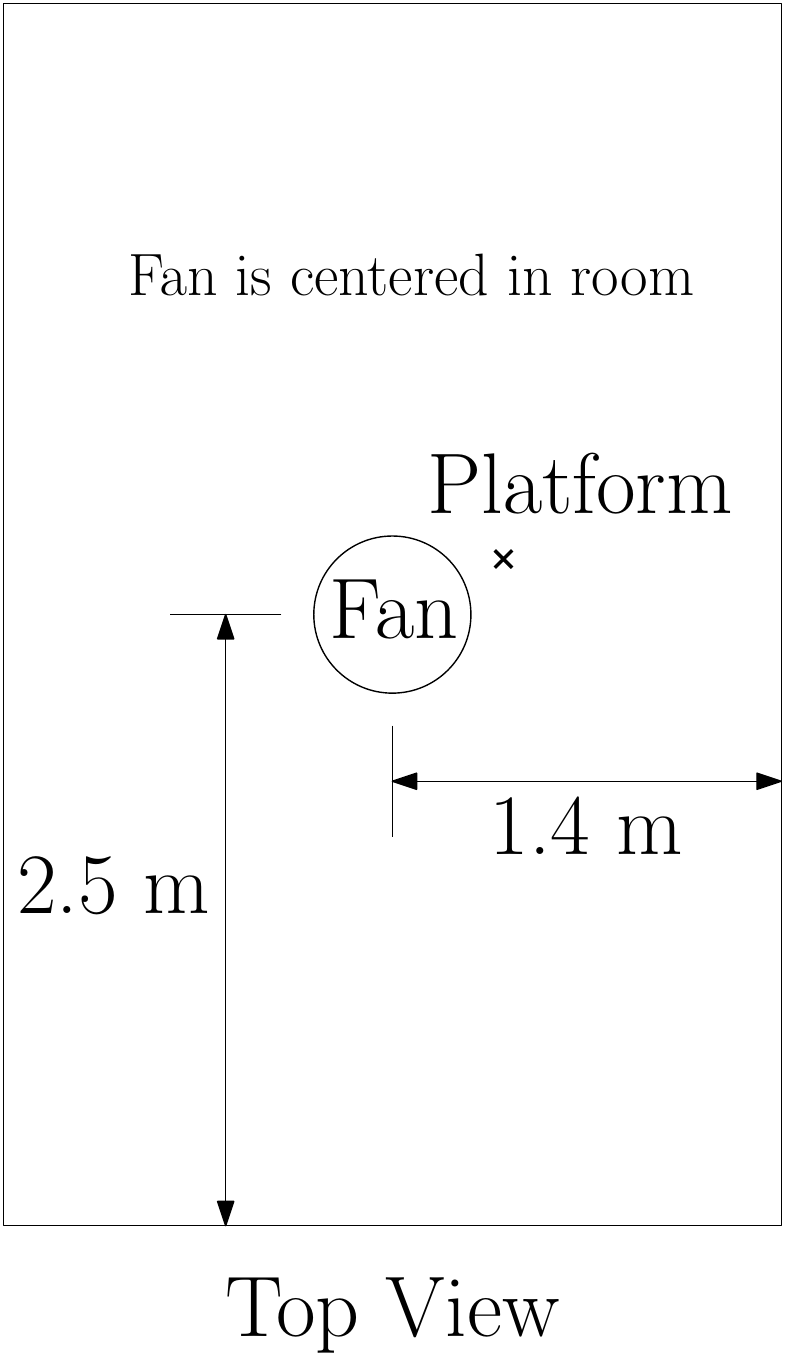}
\includegraphics[width=1.5in]{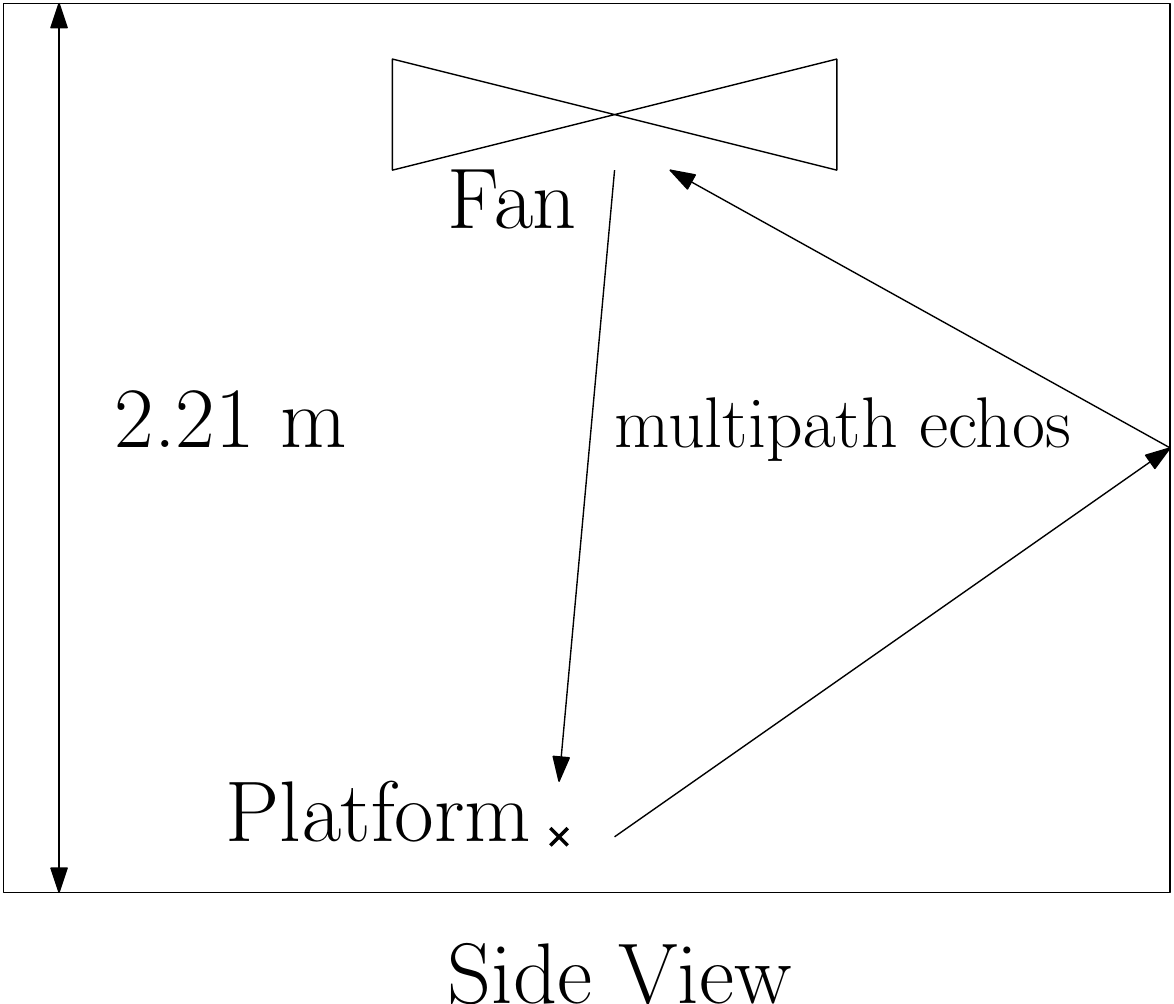}
\end{center}
\caption{Experimental floorplan}
\label{floorplan_fig}
\end{figure}

\begin{figure}
\begin{center}
\includegraphics[width=3in]{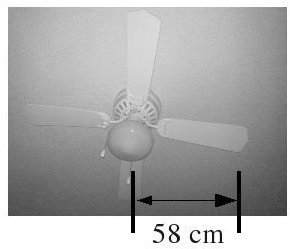}
\end{center}
\caption{The fan used in the physical experiments}
\label{target_fig}
\end{figure}

We conducted our experiments in a relatively simple geometry, as shown in Figure \ref{floorplan_fig}.  The rotating target was a ceiling fan (see Figure \ref{target_fig}) mounted on the ceiling of a small room with four walls.  The fan had four 58 cm blades, which are inclined at an angle of 15 degrees with respect to the horizontal.  To minimize acoustic noise from the motor, the fan was spun by giving it a push by hand before collecting data.

A simple ray tracing approach was used to produce the simulation dataset, with not more than one specular wall bounce permitted.  The blades were simulated as perfectly reflective line segments, which were static for the duration of each pulse.  Reflections off the simulated blades were permitted under the condition that the angles made by the incident and reflected rays and the line segment were equal.  After each pulse, the position of each of the blades was updated according to the rotation speeds observed in the physical experiment (a rotation rate of approximately 0.5 - 3 Hz).  No noise or systematic errors were included in the simulation, and the reflection coefficients for the walls were artificially large (0.5), simply because the primary output of the simulation was {\it timing information} and not signal levels.

The physical experiment was conducted in a room of the same dimensions as shown in Figure \ref{floorplan_fig}.  Each of the walls and the ceiling was composed of residential drywall, and the floor was composed of wood.  There were numerous obstacles of various acoustic properties present in the scene.  Although these obstacles were uncontrolled, they were static and therefore of relatively little importance.  

\begin{table}
\caption{Summary of Physical Collections}
\begin{center}
\begin{tabular}{c|c|c|c|c}
Collection&Direction&Fan rate&Doppler frequency&1/4 Doppler\\
&&(visual)&of vertical band&frequency\\
\hline
A&Static&0 Hz&N/A&N/A\\ % 20111008 floor
B&Static&0 Hz&N/A&N/A\\ % 2012022 static number 2
C&CCW&1/2 Hz&3.39 Hz&0.85 Hz\\ % 20111008 floor
D&CCW&1/3 Hz&1.56 Hz&0.39 Hz\\ % 20120222
E&CCW&1/2 Hz&2.34 Hz&0.58 Hz\\ % 20120222
F&CW&1/2 Hz&2.145 Hz&0.54 Hz\\ % 20120222
\end{tabular}
\end{center}
\label{speed_tab}
\end{table}

Six physical collections were performed, as summarized in Table \ref{speed_tab}.  The collection system was a monostatic sonar platform running in software on a Nokia n900 smartphone.  Two threads of execution were used: 
\begin{enumerate}
\item A transmit process, which played the desired waveform stored as a WAV file over the speakers, and
\item A receive process, which recorded the transmitted signal and subsequent echos.
\end{enumerate}
Subsequent analysis of the collected data was performed using custom-made scripts in GNU Octave \cite{octave_ref}.

\subsection{Processing methodology}
\label{proc_sec}

Both the simulation and the physical experiments used the same sonar waveform, so as to ensure comparable results.  These waveform parameters are summarized in Table \ref{sonar_tab}.

\begin{table}
\caption{Sonar collection paramters}
\begin{center}
\begin{tabular}{c|c}
Parameter&Value\\
\hline
Pulse type&Narrow impulse\\
Pulse repetition frequency&34 Hz\\
Receive sampling rate&44.1 kHz\\
Pulses collected&175\\
\end{tabular}
\end{center}
\label{sonar_tab}
\end{table}

\begin{figure}
\begin{center}
\includegraphics[width=2in]{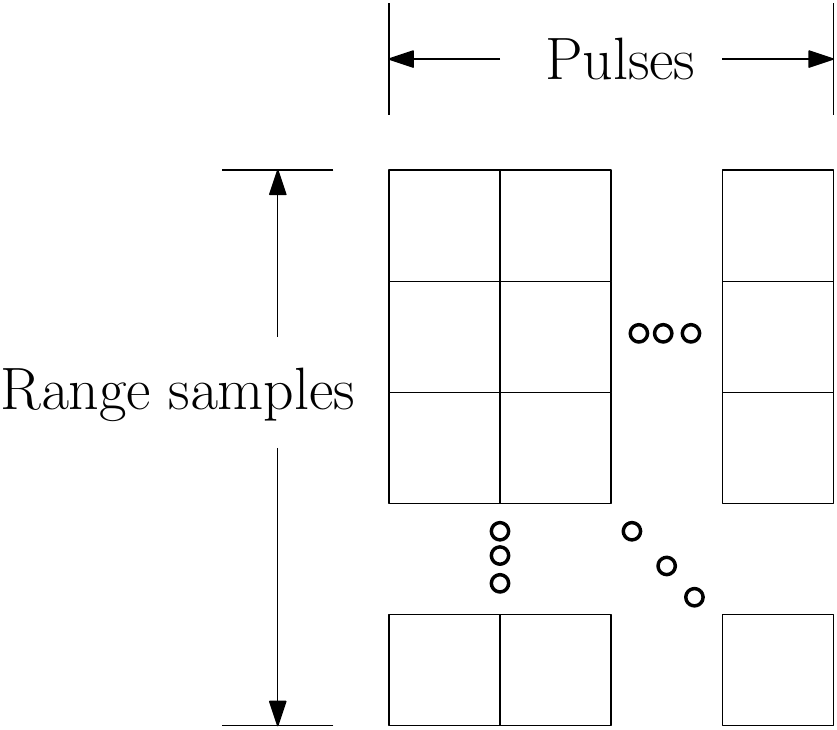}
\end{center}
\caption{Layout of the data before processing}
\label{datamatrix_fig}
\end{figure}

In both the simulation and the physical experiment, the collected data is in the form of a single contiguous collection of time samples.  These were grouped into pulses by acquiring the leading edge of the first pulse and using the known pulse interval.  The resulting data was formed into a matrix (see Figure \ref{datamatrix_fig}), whose columns correspond to the pulses and whose rows correspond to samples within each pulse.

The data was processed using three different methods: range-doppler processing, inverse synthetic aperture processing, and a novel angular fingerprinting approach.

The range-doppler processing used was quite simple: the data matrix was discrete Fourier transformed in the column (pulse) direction, and then centered at the zero doppler frequency.  No further filtering or weighting was applied, as high-resolution range-doppler results were not desired.  

We also applied an inverse synthetic aperture process to the data in order to validate the quality of the direct path (transmitter-target-receiver) data in both experiments.  We used a time-domain backprojection image formation technique \cite{Jakowatz}, projecting the data to a reference frame rotating with the target.  This is easy in simulation; the center of the fan and its rotation rate are known precisely.  However, to avoid blurring effects, this required careful measurement of the fan's rotation rate (not its direction) for the physical experiment.  Of course, using only the direct path signal for this image formation process means that no multipath effects are included.  Therefore, since the center of the fan remains fixed, there is an unavoidable ambiguity in its rotation.  For this reason, the formation of the image is insensitive to rotation direction.  (Realize that in the usual case of inverse synthetic aperture processing, the target centroid is also in motion; this destroys the ambiguity.) 

\label{finger_alg_sec}

Finally, we applied a novel angular fingerprinting process to detect changes in rotation direction and rate.  Schematically, this process relies on two collections: a reference collection, and a second collection of interest.  The resulting output indicates whether the two collections had the same or different rotation directions, and how the relative rotation rate varied over the second collection.  

Crucially, our fingerprinting algorithm relies on the injectivity of the signal profile, which is a consequence of Theorem \ref{inj_thm}.  This implies that each angular position corresponds to a {\it unique} collection of multipath signals.  We use the multipath signals acquired from the reference collection and the assumption that the speed of rotation is constant to set up a correspondence between the multipath signals acquired in the second collection and the angular position of the rotating target in the reference collection.

Precisely, the position of the fan is a smooth function of time
\begin{equation*}
P_i:\mathbb{R}\to S^1,
\end{equation*}
where $i=1$ for the reference collection and $i=2$ for the second collection.  The position of the fan gives rise to particular multipath signals, described by the signal profile $T:S^1\to(\mathbb{R}\,\sqcup\perp)^N$.  (Notice that this profile is the same for both collections.) So in particular, the measured data is a discretization of the function $M_i=T\circ P_i:\mathbb{R}\to(\mathbb{R}\,\sqcup\perp)^N$.

If $N>3$, Theorem \ref{inj_thm} argues that $T$ will generically be injective, so that $T^{-1}$ is a well-defined {\it function} on the image of $T$.  Thus, we define the {\it angular fingerprint} to be the composition of functions $F:T^{-1} \circ T \circ P_2:\mathbb{R} \to S^1$, which computes the angular position of the fan from the measurements of the second collection.  Our angular fingerprinting algorithm approximates this composition in a two step process:
\begin{enumerate}
\item Measurement of the period of rotation in the reference collection.  Under the assumption of constant speed of rotation, this results in an explicit linear regression formula for $P_1$.
\item Approximation of $T^{-1} \circ T$.  Specifically, we use $T=T \circ P_i \circ P_i^{-1}=M_i \circ P_i^{-1}$, where of course $P_i^{-1}$ is well-defined over a single period by the constant rotation speed assumption.
\end{enumerate}

Operationally, the process consists of the following steps:
\begin{enumerate}
\item Acquisition of the reference collection.
\item Measurement of the rotation period. We used a range-doppler filter to identify the target's signature, from which the rotation period can be deduced as the largest doppler component at the target's range.  Note that due to the fact that the fan had 4 blades, the measured period corresponds to an angular traversal of $\pi/2$. 
\item Range gating.  Since the target's range is known, all range bins before this range are removed.  This is important since there were substantial range and doppler sidelobes present in the waveform.  (These are largely due to receiver desense and overload effects as no time sensitivity control was applied to the receiver.)
\item Storage of a contiguous block of pulses that correspond to precisely one period.  Assuming the fan is rotating at a constant speed, the pulse number corresponds linearly with the angular coordinate of the fan's rotation.
\item Acquisition of the second collection.
\item Application of the same range gate filtering as applied to the reference collection.
\item For each pulse of the second collection, the nearest pulse (in the energy norm) of the first collection is computed.  This sets up a function from the pulses of the second collection to the angular coordinate of the first.  
\item Since this function is subject to noise, we Kalman filter it, resulting in an approximation of the angular fingerprint function $F$ is the output of the procedure.
\end{enumerate}
By examining the slope of the angular fingerprint, changes in rotation direction speed can be deduced.  If the slope is negative, then the rotation directions in the first and second collections differ; if positive, then the rotation directions agree.

\section{Analysis and implications}

\begin{figure}
\begin{center}
\includegraphics[width=3.5in]{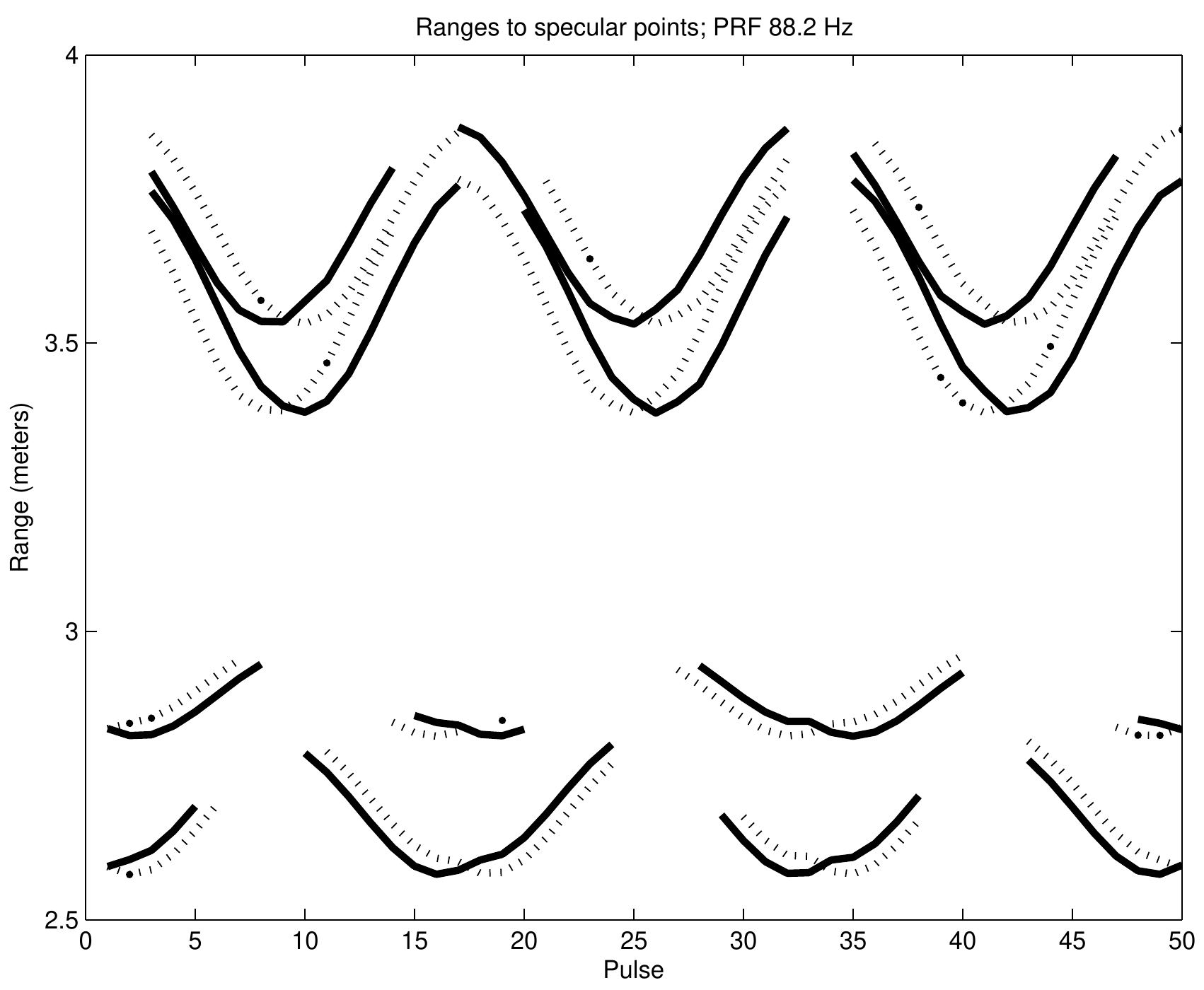}
\end{center}
\caption{Range timeseries of the simulated datasets showing only the multipath components.  Solid lines correspond to fan rotating counterclockwise; dashed lines represent clockwise rotation.}
\label{range_pri1300}
\end{figure}

The multipath components of the simulated data is shown in Figure \ref{range_pri1300}, which shows the range to specular reflection points as a function of pulse number.  Both directions of rotation are shown.  Since the plots of the two directions differ even after ignoring horizontal translations, it is clear that the direction of rotation is visible when multipath is considered.

\subsection{Range-doppler analysis}
\begin{figure}
\begin{center}
\includegraphics[width=3.5in]{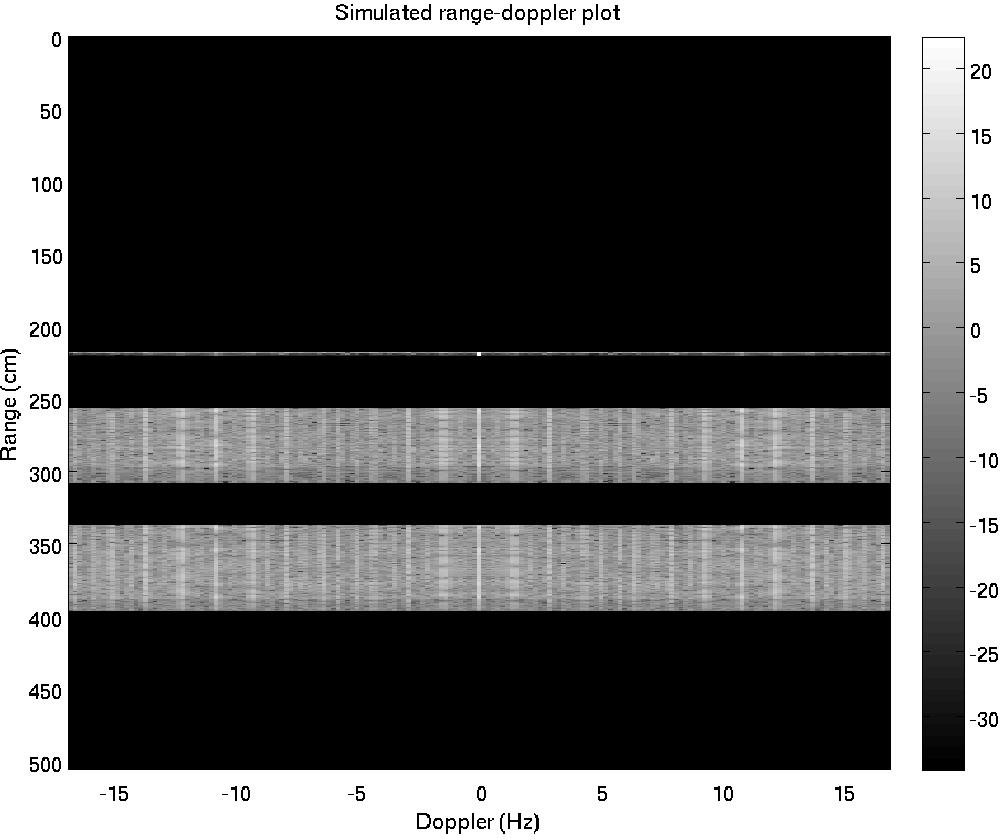}
\end{center}
\caption{Simulated range doppler plot; the direct path echo is at 219 cm, multipath echos occur in the vicinity of ranges of 250 cm - 400 cm}
\label{rd_sim_fig}
\end{figure}

\begin{figure}
\begin{center}
\includegraphics[width=3.5in]{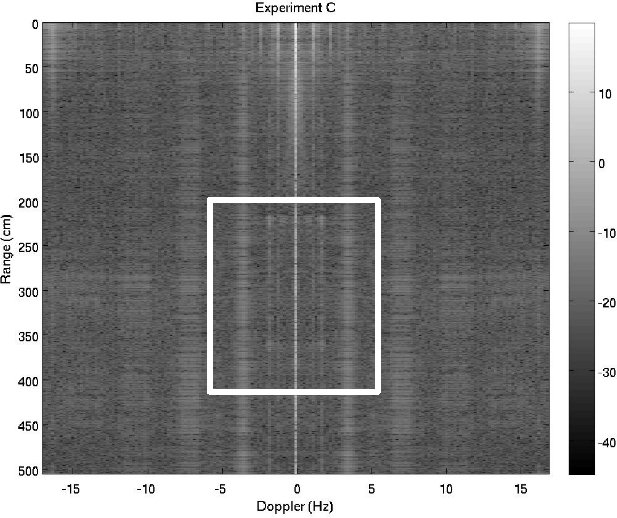}
\end{center}
\caption{Typical experimental range doppler plot (collection C); note vertical banding effect.  The marked region is shown in Figure \ref{rd_detail}.}
\label{rd_compare_fig}
\end{figure}

\begin{figure}
\begin{center}
\includegraphics[width=3.5in]{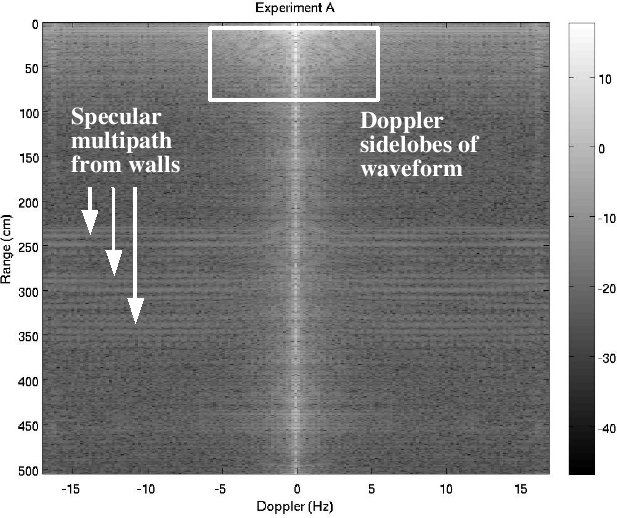}
\end{center}
\caption{Range-doppler plot of static experiment (collection A)}
\label{rd_static}
\end{figure}

The resulting range-doppler plot (obtained by discrete Fourier transforming along the pulses direction) arising from the simulated data is shown in Figure \ref{rd_sim_fig}.  In contrast, Figure \ref{rd_compare_fig} shows a collected physical experimental in which the fan was spinning.  Most of the small-range response is actually due to imperfections in the waveform, as is immediately clear from the Figure \ref{rd_static}, in which nothing in the scene was moving.

\begin{figure}
\begin{center}
\includegraphics[width=3.5in]{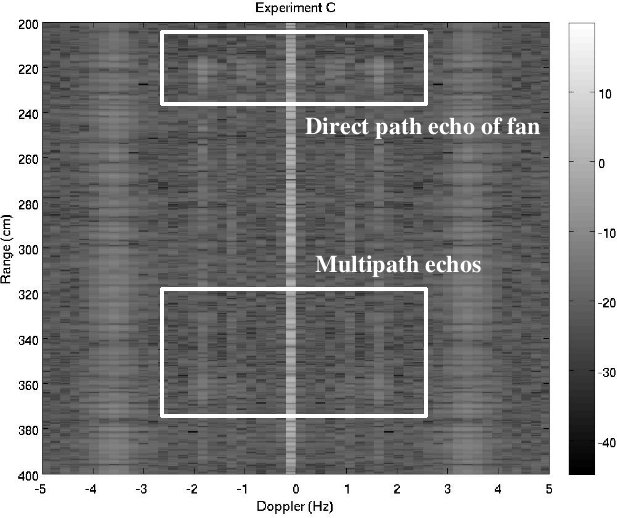}
\end{center}
\caption{Detail of experimental range doppler plots (collection C), showing micro-doppler structure of fan and multipath}
\label{rd_detail}
\end{figure}

The simulated data is quite helpful in locating the direct path and multipath responses in the physical experiments.  In particular, the direct path response is present at a range of 219 cm, and multipath responses are visible between 260-300 cm and 340-380 cm.  For instance, some multipath echos and microdoppler structure in the vicinity of the direct path response can be seen in Figure \ref{rd_detail}.

One striking difference between the simulated and physical experiments' range-doppler images is a vertical banding effect.  (This effect is also sometimes seen in windfarm time-frequency plots \cite{Isom_2009}.)  The physical experiment (Figure \ref{rd_compare_fig}) shows bright bands covering all ranges at specific doppler values, while the simulation shows no such effect.  Surprisingly, as Table \ref{speed_tab} shows, the bands are representative of the fan tip speed, even though the collection geometry is not favorable for making this measurement.  Indeed, the direct path is nearly parallel to the fan's axis of rotation, so the direct-path doppler is quite low.  Thus, it is quite likely an effect of multipath signals, for which the expected doppler can be higher.  Indeed, the simulated result (Figure \ref{rd_sim_fig}) shows stronger, doppler-localized responses within the ranges associated to multipath.

The fact that the banding effect covers all ranges indicates that it is in fact an effect that extends in time longer than a pulse interval, and is not divisible by a pulse interval.  This suggests that it is not an effect of the propagating ray paths and their single-bounce path lengths; these are all captured by the simulation.  However, it is likely that some multi-bounce path could be nearly specular.  Such a path could therefore return acoustic energy to the platform with substantial delays, which would then result in considerable range smearing.  

\subsection{Inverse synthetic aperature analysis}
\begin{figure}
\begin{center}
\includegraphics[width=1.5in]{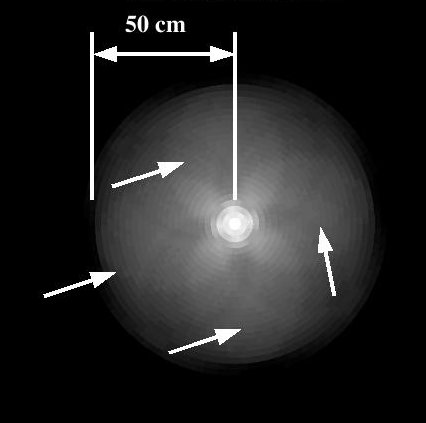}
\includegraphics[width=1.5in]{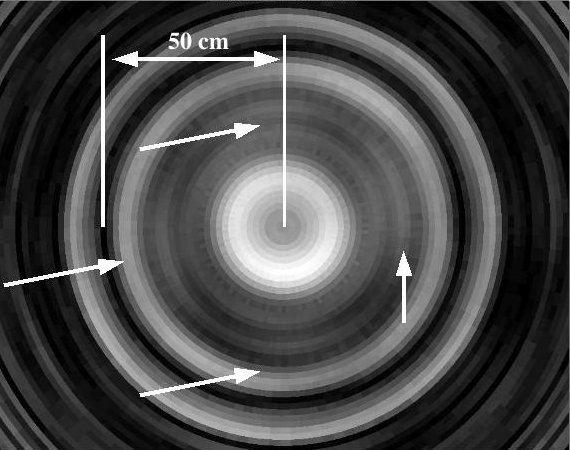}
\end{center}
\caption{Comparison between simulated (left) and experimental (right) inverse synthetic aperature images. Note the presence of four blades (marked with arrows)}
\label{isar_compare_fig}
\end{figure}

Range-doppler analysis is not a precise way to analyze the quality of the direct path echos.  The formation of an inverse synthetic aperature image is considerably more demanding of timing stability and location accuracy.  Both the simulation and physical experiments indeed have sufficient resolution to correctly resolve the blades of the fan, as can be seen in Figure \ref{isar_compare_fig}.  There is also a bright circular response around the {\it outside} of the fan in the physical experiment's image.  This could be a delayed re-radiation due to vibrational resonance in the fan assembly, but this is not conclusive.

Of course as described elsewhere in this article, direction of rotation cannot be discerned from direct path measurements.  Switching the rotation direction of the image plane results in nearly identical images (after accounting for a random rotation).
 
\subsection{Angular fingerprinting}
\label{fingerprint_sec}
The angular fingerprinting algorithm was applied to each pair of physical experiments.  Because of the strong doppler sidelobes present in the waveform, the responses from the first 250 samples were ignored, which corresponds to a range of 194 cm.  Since the range to the target was 216 cm, this preserved the direct path response.  After a preliminary angular fingerprint function was computed, we smoothed the data with a simple Kalman filter with a maximum slew rate of 4 pulses per pulse.

\begin{figure}
\begin{center}
\includegraphics[width=3.5in]{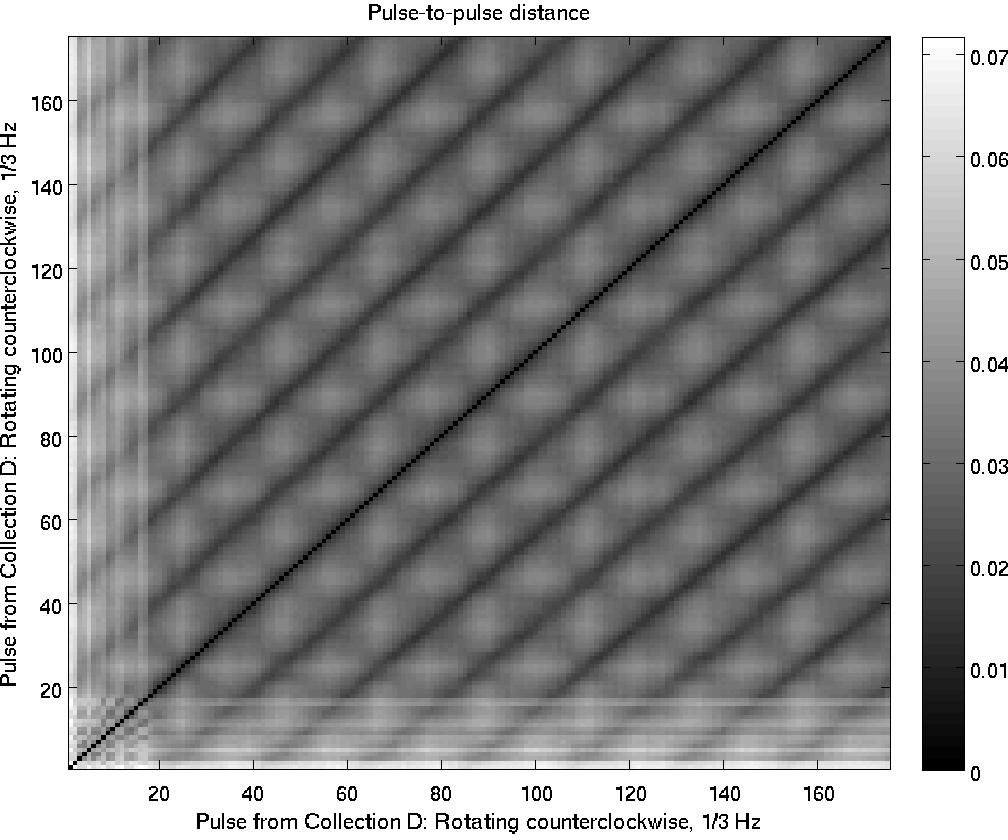}
\end{center}
\caption{Pairwise matching distance between pulses of collection D}
\label{self_fingerprint_fig}
\end{figure}

\begin{figure}
\begin{center}
\includegraphics[width=3.5in]{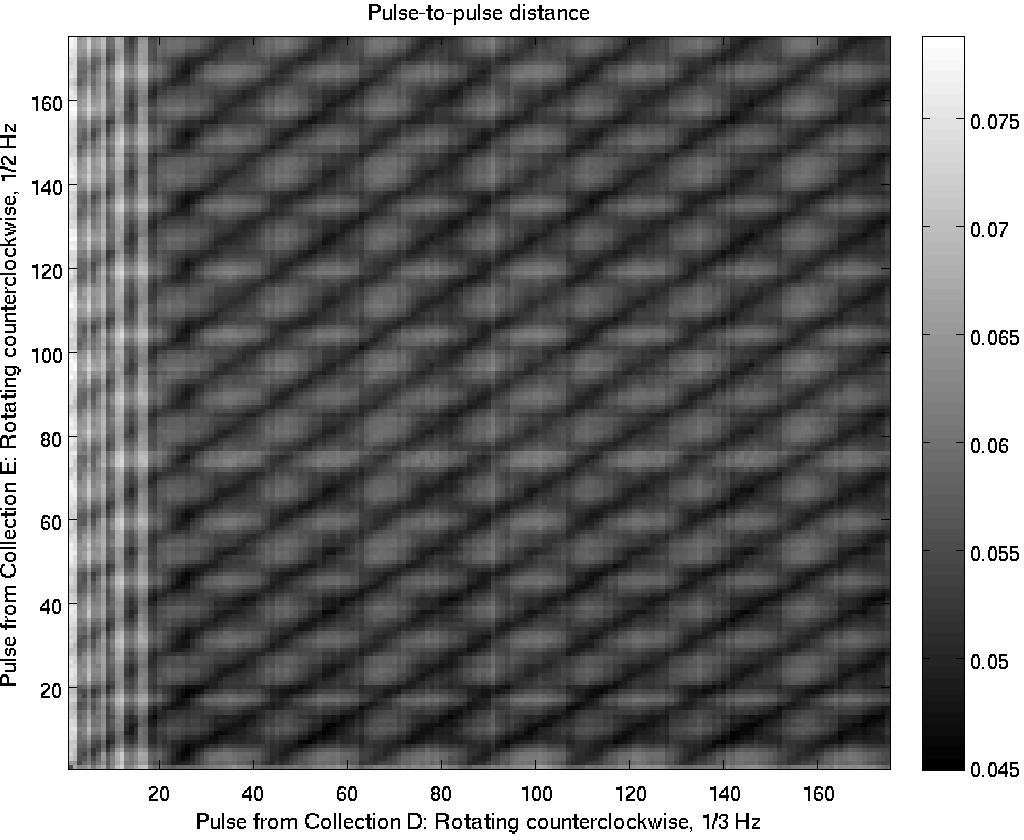}
\end{center}
\caption{Pairwise pulse-to-pulse matching distance of two experiments with the same rotation direction (collections D and E)}
\label{fingerprint1_fig}
\end{figure}

\begin{figure}
\begin{center}
\includegraphics[width=3.5in]{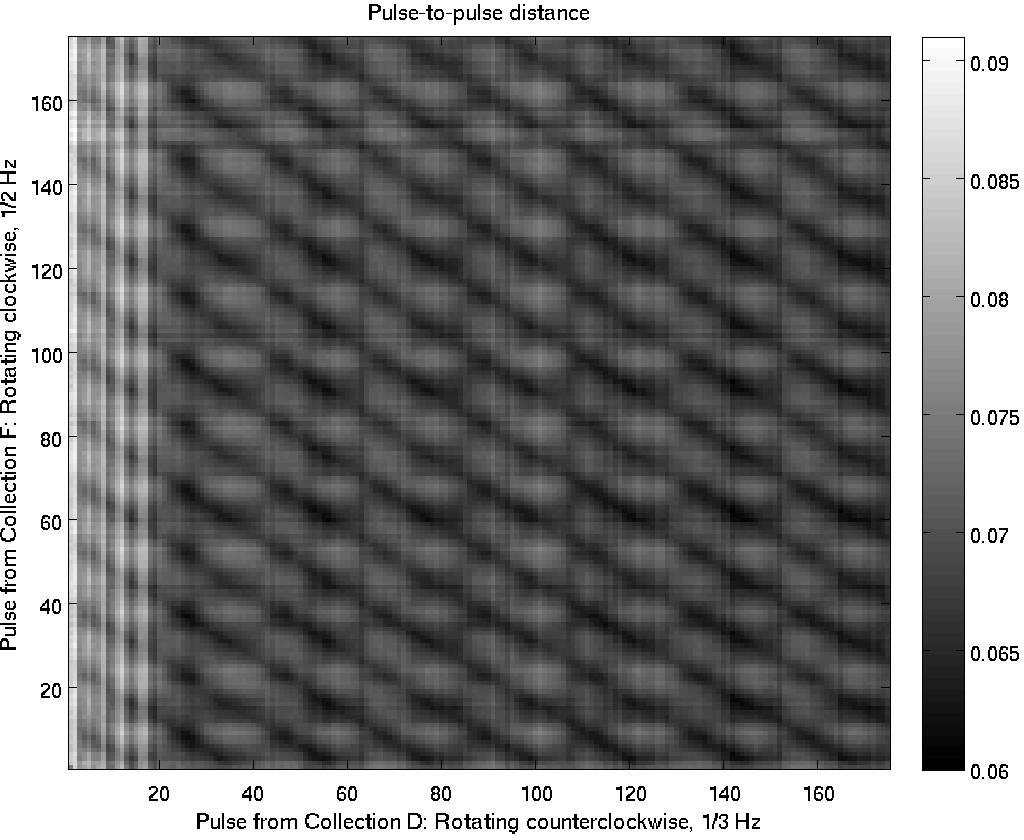}
\end{center}
\caption{Pairwise pulse-to-pulse matching distance of two experiments with different rotation directions and speeds (collections D and F)}
\label{fingerprint2_fig}
\end{figure}

\begin{figure}
\begin{center}
\includegraphics[width=3.5in]{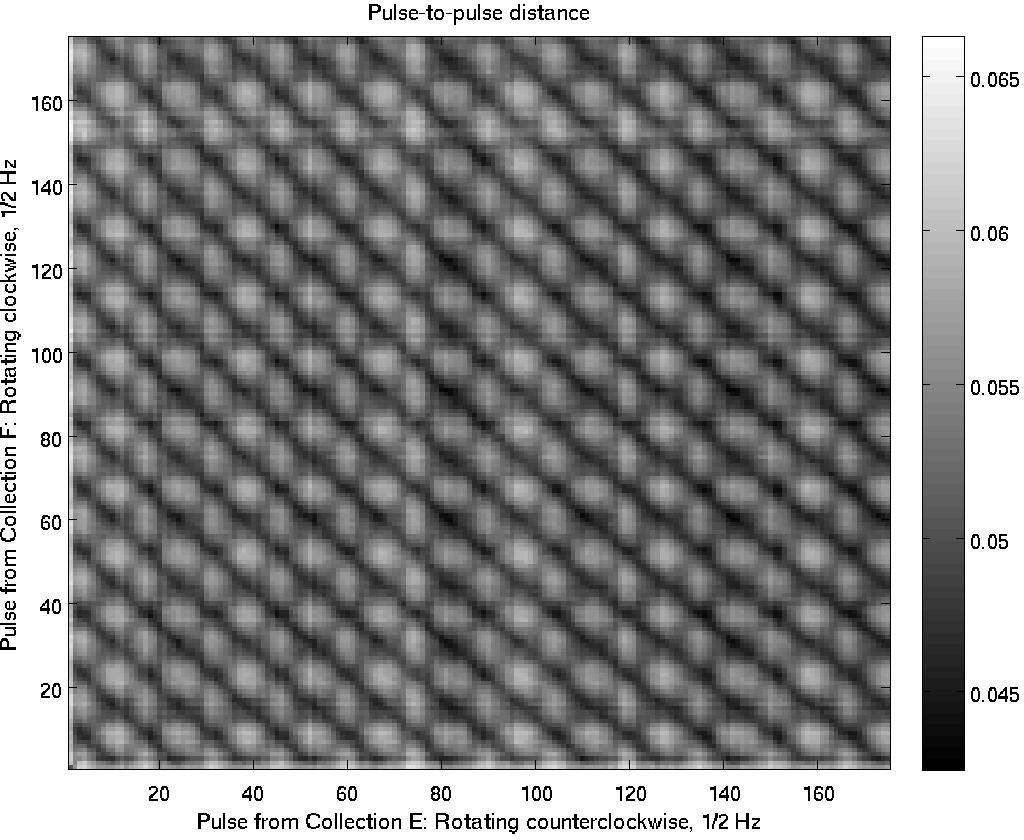}
\end{center}
\caption{Pairwise pulse-to-pulse matching distance of two experiments with different rotation directions but similar speed (collections E and F)}
\label{fingerprint3_fig}
\end{figure}

\begin{figure}
\begin{center}
\includegraphics[width=3.5in]{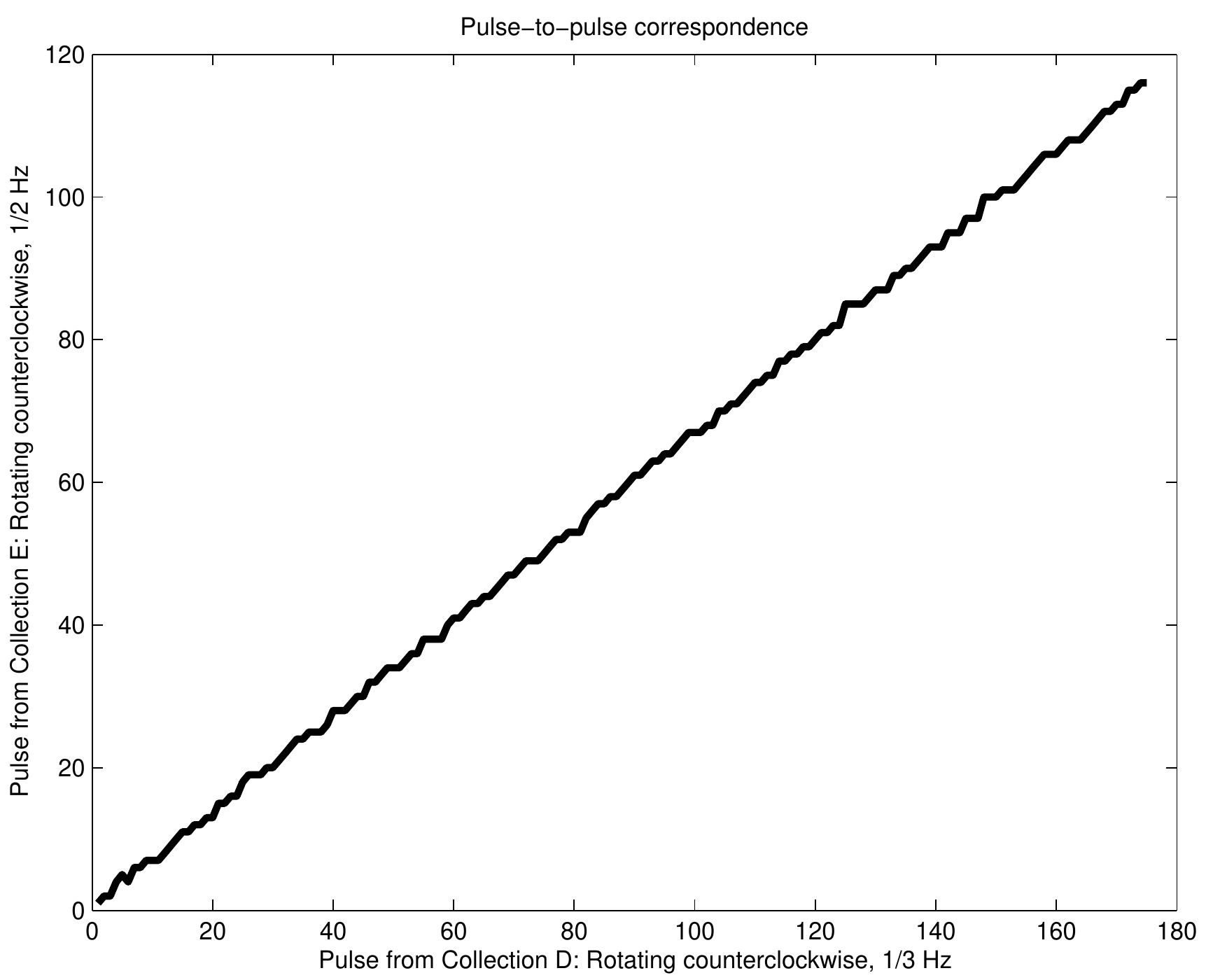}
\end{center}
\caption{Angular fingerprint of two experiments with the same rotation direction (collections D and E)}
\label{filtered1_fig}
\end{figure}

\begin{figure}
\begin{center}
\includegraphics[width=3.5in]{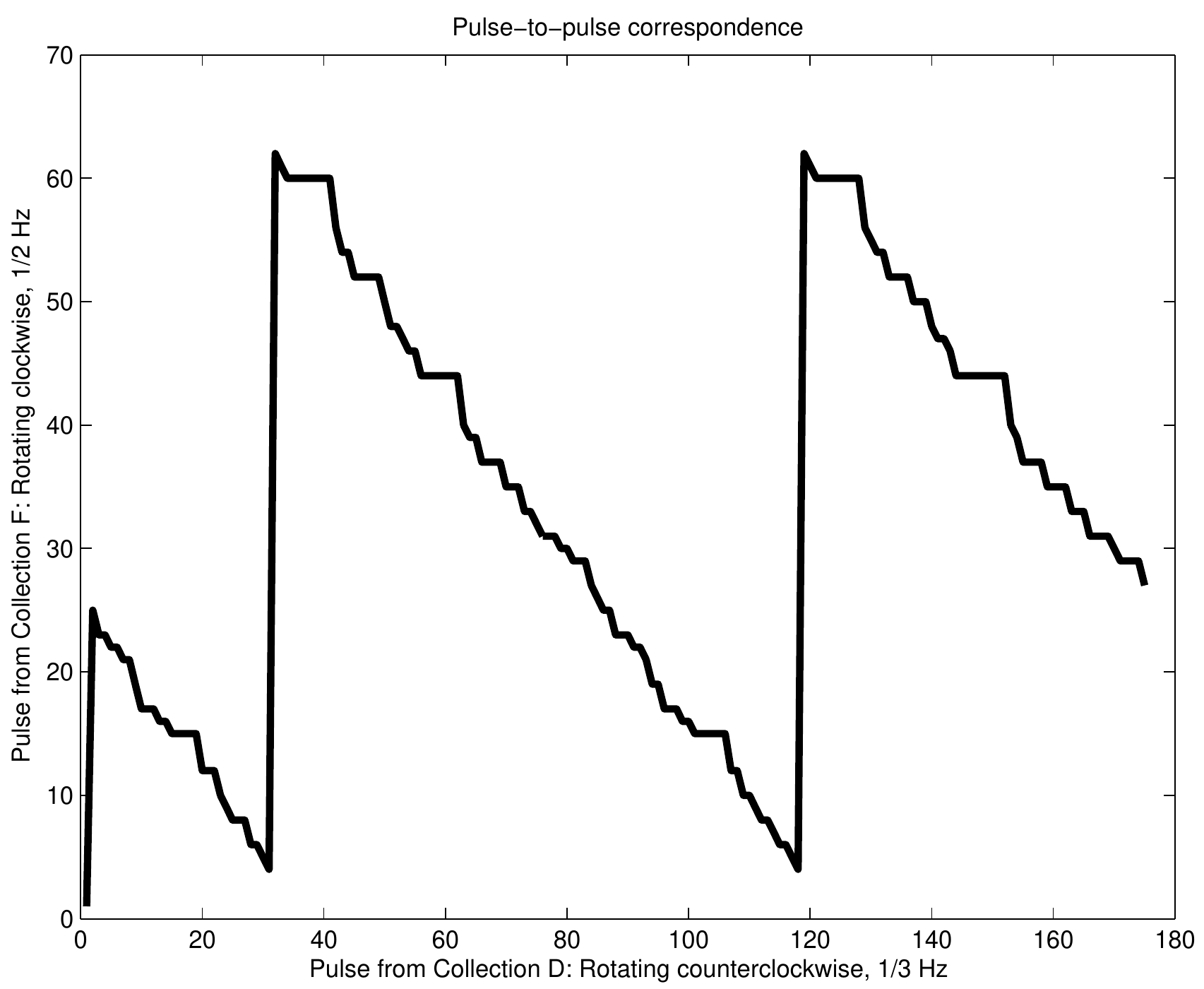}
\end{center}
\caption{Angular fingerprint of two experiments with different rotation directions and speeds (collections D and F)}
\label{filtered2_fig}
\end{figure}

\begin{figure}
\begin{center}
\includegraphics[width=3.5in]{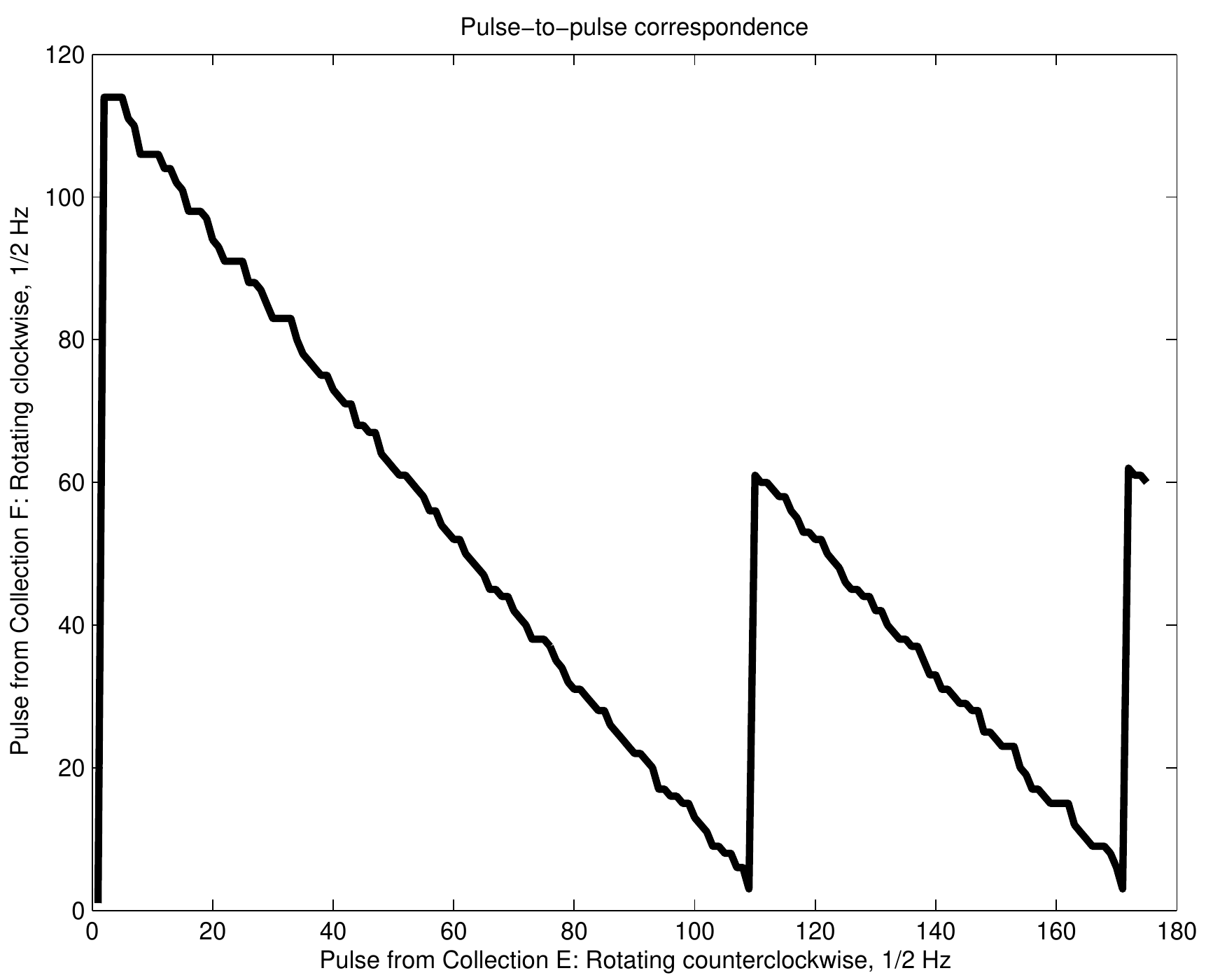}
\end{center}
\caption{Angular fingerprint of two experiments with different rotation directions but similar speed (collections E and F)}
\label{filtered3_fig}
\end{figure}

Sample results of the angular fingerprinting process are shown in Figures \ref{self_fingerprint_fig}-\ref{filtered3_fig}.  In Figure \ref{self_fingerprint_fig}, the best case scenario is shown; the reference and second collections are identical, so there should be a perfect match between each pulse and its copies at subsequent periods.  Indeed, the periodic structure is quite prominent.
 
More interesting are the results shown in Figures \ref{fingerprint1_fig}-\ref{fingerprint3_fig}.  These compare several different collections, in which rotation speed and direction have changed.  When the direction of rotation has changed, the slope of the diagonal stripes has changed sign.  When the speed is different, the magnitude of the slope differs from unity.  These results agree quite well with the collection truth data in Table \ref{speed_tab}.

The distance plots in Figures \ref{fingerprint1_fig}-\ref{fingerprint3_fig} are an intermediate step in the angular fingerprint process.  The final output fingerprint functions (after Kalman filtering) are displayed in Figures \ref{filtered1_fig}-\ref{filtered3_fig}.  Again, the slope of these functions match up quite closely with the truth.

\section{Conclusion and future directions}
The experiments presented in this article show that doppler multipath is a valuable source of signal for making observations that would be impossible otherwise.  While exploitation of multipath has been difficult using methods that require substantial knowledge of the environment, a topological approach reveals that one can sometimes remove this requirement entirely.

With more substantial knowledge of the environment, it is possible that this approach can be improved.  Indeed, if we take the ideas of \cite{Tayebi_2009} seriously, then we may be able to use a {\it simulated} reference collection instead of taking physical measurements.  This requires a very detailed multipath model, with a close correspondence between the geometry of the simulation and the physical experiment.  The simple multipath model presented in this article is insufficiently detailed for the purpose of supplying such a reference collection.  Our attempts in this direction have been unsuccessful, however this appears to be an interesting direction for future efforts.

\section*{Acknowledgment}
This work was supported by AFOSR FA9550-09-1-0643.
%\small

%\normalsize

\bibliographystyle{IEEEtran}
\bibliography{ceilingfan_bib}

\begin{thebibliography}{10}
\providecommand{\url}[1]{#1}
\csname url@samestyle\endcsname
\providecommand{\newblock}{\relax}
\providecommand{\bibinfo}[2]{#2}
\providecommand{\BIBentrySTDinterwordspacing}{\spaceskip=0pt\relax}
\providecommand{\BIBentryALTinterwordstretchfactor}{4}
\providecommand{\BIBentryALTinterwordspacing}{\spaceskip=\fontdimen2\font plus
\BIBentryALTinterwordstretchfactor\fontdimen3\font minus
  \fontdimen4\font\relax}
\providecommand{\BIBforeignlanguage}[2]{{%
\expandafter\ifx\csname l@#1\endcsname\relax
\typeout{** WARNING: IEEEtran.bst: No hyphenation pattern has been}%
\typeout{** loaded for the language `#1'. Using the pattern for}%
\typeout{** the default language instead.}%
\else
\language=\csname l@#1\endcsname
\fi
#2}}
\providecommand{\BIBdecl}{\relax}
\BIBdecl

\bibitem{Poupart_2003}
G.~Poupart, ``Wind farms impact on radar aviation interests,'' Qinetiq, Tech.
  Rep. DTI PUB URN 03/1294, 2003.

\bibitem{Tayebi_2009}
A.~Tayebi, J.~Gomez, F.~S. de~Adana, and O.~Gutierrez, ``The application of
  ray-tracing to mobile localization using the direction of arrival and
  received signal strength in multipath indoor environments,'' \emph{Progress
  in Electromagnetics Research}, vol. PIER 91, pp. 1--15, 2009.

\bibitem{GhristRobinson}
R.~Ghrist and M.~Robinson, ``Topological localization via signals of
  opportunity,'' \emph{{IEEE} Trans. Sig. Proc}, 2012.

\bibitem{Perry_2007}
J.~Perry and A.~Biss, ``Wind farm clutter mitigation in air surveillence
  radar,'' \emph{IEEE Aerospace and Electronics Magazine}, vol.~22, no.~7, July
  2007.

\bibitem{Isom_2009}
B.~M. Isom, R.~D. Palmer, G.~S. Secrest, R.~D. Rhoton, D.~Saxion, T.~L. Allmon,
  J.~Reed, T.~Crum, and R.~Vogt, ``Detailed observations of wind turbine
  clutter with scanning weather radars,'' \emph{Journal of Atmospheric and
  Oceanic Technology}, vol.~26, May 2009.

\bibitem{Chen_2001}
V.~C. Chen and W.~J. Miceli, ``Simulation of {ISAR} imaging of moving
  targets,'' \emph{Radar, Sonar, \& Navigation, IET}, vol. 148, no.~3, pp.
  160--166, June 2001.

\bibitem{chen_2006}
V.~Chen, F.~Li, S.-S. Ho, and H.~Weschler, ``Micro-doppler effect in radar:
  phenomenon, model, and simulation study,'' \emph{IEEE Transactions on
  aerospace and electronic systems}, vol.~42, no.~1, January 2006.

\bibitem{Thayaparan_2006}
T.~Thayaparan, ``Separation of target rigid body and microdoppler effects in
  {ISAR/SAR} imaging,'' DRDC Ottawa, Tech. Rep. TM 2006-187, September 2006.

\bibitem{Thayaparan_2007}
T.~Thayaparan, S.~Abrol, E. Riseborough, L.~Stankovic, D.~Lamothe, and
  G.~Duff, ``Analysis of radar micro-doppler signatures from experimental
  helicopter and human data,'' \emph{Radar, Sonar, \& Navigation, IET}, vol.~1,
  no.~4, August 2007.

\bibitem{Suwa_2009}
K.~Suwa, T.~Wakayama, and M.~Iwamoto, ``Estimation of target motion and 3d
  target geometry using multistatic {ISAR} movies,'' in \emph{IGARSS}, July
  2009.

\bibitem{Sherman_1971}
S.~M. Sherman, ``Complex indicated angles applied to unresolved targets and
  multipath,'' \emph{IEEE Trans. Aerospace Electron Syst.}, vol. AES-7, no.~1,
  January 1971.

\bibitem{Hahm_1997}
M.~D. Hahm, Z.~I. Mitrovski, and E.~L. Titlebaum, ``Deconvolution in the
  presence of doppler with application to specular multipath parameter
  estimation,'' \emph{IEEE Trans. Sig. Proc.}, vol.~45, no.~9, September 1997.

\bibitem{Kelly_2000}
I.~Kelly, H.~Deng, and H.~Ling, ``On the feasibility of the multipath
  fingerprint method for location finding in urban environments,''
  \emph{Applied Computational Electromagnetics Society Journal}, vol.~15,
  no.~3, pp. 232--247, November 2000.

\bibitem{Prasith_2002}
P.~Prasithsangaree and P.~Krishnamurthy, ``On indoor position location with
  wireless {LANs},'' in \emph{The 13th IEEE Int. Conf. on Personal, Indoor, and
  Mobile Radio Communications}, 2002.

\bibitem{Nerguizian_2006}
C.~Nerguizian, C.~Despins, and S.~Aff\`es, ``Geolocation in mines with an
  impulse response fingerprinting technique and neural networks,'' \emph{IEEE
  Transactions on Wireless Communications}, vol.~5, no.~3, March 2006.

\bibitem{Roxin_2007}
A.~Roxin, J.~Gaber, M.~Wack, and A.~Nait-Sidi-Moh, ``Survey of wireless
  geolocation techniques,'' in \emph{IEEE Globecom Workshops}, 2007.

\bibitem{Fang_2008}
S.-H. Fang, T.-N. Lin, and K.-C. Lee, ``A novel algorithm for multipath
  fingerprinting in indoor {WLAN} environments,'' \emph{IEEE Trans. Wireless
  Communications}, vol.~7, no.~9, September 2008.

\bibitem{Charvat_2006}
G.~Charvat, ``Low-cost, high-resolution laboratory radar system for synthetic
  aperature radar applications,'' in \emph{AMTA}, October 2006.

\bibitem{Matejowsky_2008}
E.~Matejowsky, ``Home made sonar, {\tt
  http://eddiem.com/projects/chirp/chirp.htm}.''

\bibitem{Laan_2009}
L.~Labs, ``Sonar ruler {iPhone} app,'' 2009.

\bibitem{Dicon_2012}
DiCon, ``Sonar android app,'' January 2012.

\bibitem{octave_ref}
J.~W. Eaton, D.~Bateman, and S.~Hauberg, \emph{GNU Octave Manual Version
  3}.\hskip 1em plus 0.5em minus 0.4em\relax Network Theory Limited, 2008.

\bibitem{Jakowatz}
C.~Jakowatz, D.~Wahl, P.~Eichel, and P.~Thompson, \emph{Spotlight mode
  synthetic aperature radar: a signal processing approach}.\hskip 1em plus
  0.5em minus 0.4em\relax Springer, 1996.

\end{thebibliography}

%\begin{biography}[{\includegraphics[width=1in,height=1.25in,clip,keepaspectratio]{mshell}}]{Michael Shell}

\begin{IEEEbiography}
[{\includegraphics[width=1in]{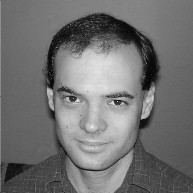}}] {Michael
  Robinson} is a postdoctoral fellow in the Department of Mathematics
at the University of Pennsylvania and a research engineer at SRC,
inc. His 2008 Ph.D. in Applied Mathematics [Cornell University] and
recent work in topological signal processing is complemented by a
background in Electrical Engineering and current work in radar systems
analysis.
\end{IEEEbiography}
\end{document}